\documentclass[fleqn,twoside]{article}
\usepackage{espcrc2}

\usepackage{graphicx,epsfig}
\usepackage[figuresright]{rotating}

\def \gto{\stackrel{\gamma}{\to}}
\def \bra#1{\langle #1 |}
\def \ket#1{| #1 \rangle}
\def \br{${\cal B}$}

\newcommand{\AmS}{{\protect\the\textfont2
  A\kern-.1667em\lower.5ex\hbox{M}\kern-.125emS}}

\title{
\vskip -15mm
{\hfill \small DESY 02-170} \\
\vskip 5mm
Topics in Meson Spectroscopy \thanks{This work was partially funded by the Natural Sciences 
and Engineering Research Council of Canada.}}

\author{Stephen Godfrey\address{Department of Physics,
Carleton University, Ottawa K1S 5B6 CANADA} %\\
\address{DESY, Deutsches Elektronen-Synchrotron, D22603 Hamburg, 
GERMANY}%
}
       
\begin{document}

\begin{abstract}
In this mini-review I discuss three topics in meson spectroscopy.  The 
production of heavy quarkonium states, S-wave scattering below 1~GeV, 
and exotic hybrid meson production. This is not intended to be a 
comprehensive review, just an overview of several topics of current 
interest.
\vspace{1pc}
\end{abstract}

% typeset front matter (including abstract)
\maketitle

\section{INTRODUCTION}

Meson spectroscopy is an extremely broad subject, covering $\pi\pi$ 
scattering with $\sqrt{s}< 1$~GeV to the heaviest $b\bar{b}$ states 
with masses greater than 10~GeV \cite{gn99,gi85}. 
The goal at both ends of the energy scale 
is to understand QCD in the nonperturbative regime.
It is impossible to do justice to 
such a large body of work in a short talk  
so I restrict myself to a few selected topics.
Many other aspects of meson spectroscopy are covered in other 
contributions where the interested reader can 
obtain a more balanced view of recent developments.
% than is presented here.

In a very short period of time, 
the subject of heavy quarkonium has come to the 
forefront.  This is the result of two recent developments; the 
CLEO collaboration is taking data at the $\Upsilon(3S)$ and 
$\Upsilon(2S)$ which will exceed the world's existing data by a 
large factor and the B-factories have 
observed a number of charmonium states in B-decay 
\cite{belle02,belle-chi,babar-chi} culminating in the discovery of 
the $\eta_c'$ in $B\to \eta_c' K$ by the Belle 
collaboration \cite{belle02}.  
These measurements have led to the hope that some 
of the missing charmonium states may also be observed in B 
decay.  Simultaneously, there has been great progress in 
Lattice QCD \cite{lattice} so that we have progress on both the 
experimental and theoretical fronts.
%In the next section I will discuss search strategies 
%that can be used to find some of the missing quarkonium states.

On a second front there continues to be vigorous discussion about the 
structure seen in $\pi\pi$ S-wave scattering below 1~GeV.  
My impression is that a consensus is emerging among workers in the 
field with respect to general features.  However, there continues to be 
considerable debate about the details. Low energy 
S-wave scattering is discussed in Section 3.

Finally, one of the most important topics in meson spectroscopy is the 
issue of gluonic excitations.  To highlight this topic I briefly 
describe a 
proposal that should make a definitive statement about the existence 
of these objects and map out their properties.

\section{HEAVY QUARKONIUM}

The discovery of a D-wave 
$b\bar{b}$ state by the CLEO collaboration \cite{cleo02} 
and the discovery of the $\eta_c'$ in B-decay 
by the BELLE collaboration \cite{belle02} represent important advances 
in heavy quarkonium physics.
With continued running by CESR 
on the $\Upsilon(2S)$ and $\Upsilon(3S)$, large samples of 
B-mesons being collected by the B-factories, and future measurements 
at CLEO-c/CESR-c, I anticipate further 
discoveries in the near future.  At the same time,
advances in Lattice QCD calculations of the heavy 
quarkonium spectra are leading to quantitative predictions which need 
to be tested against experiment \cite{lattice}.  
Taken together we should expect a 
much better understanding of heavy quarkonium from first principles QCD.

Numerous $b\bar{b}$ states are expected to exist below 
the $B\bar{B}$ decay threshold.  However, only spin triplet 
states have been observed in the $b\bar{b}$ system and only 2 spin-singlet 
states have been observed in $c\bar{c}$.  Given that there is a wide 
variation of the spin dependent splittings predicted by various 
calculations,
the observation of additional quarkonium states, including the missing
spin-singlet states, would pose an 
important test of the calculations
 (comparisons are given in Ref. \cite{gr-dw,gr-etab,gr-hb}).  
Similarly, various 
calculations predict different values for the mass of $b\bar{b}$ D-wave cog 
(the spin weighted mass of the triplet states).
% as shown in Fig. 1.  
Observation of these states would be a powerful discriminator of the 
various calculations.

CESR/CLEO has just completed a high statistics data taking run at the 
$\Upsilon(3S)$ and will soon be running at the $\Upsilon(2S)$.  
We expect that analysis of the data will yield a rich spectroscopy.  
To give direction to searches we use the quark model to 
estimate radiative widths and branching ratios.

\subsection{D-Wave States}

There are several strategies to search for D-wave mesons \cite{gr-dw}.  
The first is via direct scans in $e^+e^-$ collisions to produce the 
$^3D_1$.  However, given the small coupling of the $^3D_1(b\bar{b})$ 
to the photon and the effects of line smearing this approach would 
require substantial statistics.  A second approach is via the
electromagnetic cascades $\Upsilon(3S) \to \gamma \chi_b' \to \gamma 
\gamma ^3D_J$.  Using quark model estimates of both E1 dipole 
transitions and decays to final state hadrons \cite{KR} we 
estimate the number of events for $4\gamma$ cascades that proceed 
via various intermediate states \cite{gr-dw}.  These are given in 
Table 1.  We expect $\sim 38$ events per $10^6$ 
$\Upsilon(3S)$'s to be produced via $^3D_J$ states.  Backgrounds can be 
reduced using the $\Upsilon \to e^+e^- + \mu^+\mu^-$ final states as 
event tags.  A substantial background will be $4\gamma$ cascades 
proceeding via the $2^3S_1$ state 
%rather than $1^3D_J$ states 
which will also produce $\sim 38$ events per $10^6$ $\Upsilon(3S)$'s.  These 
events can be separated from the $1^3D_J$ intermediate state using the 
fact that the energies of the intermediate photons are different.  The 
CLEO collaboration
successfully employed this strategy for the first observation 
of a triplet $\Upsilon(1D)$ state which was announced at this conference 
\cite{cleo02}.

\begin{table}[htb]
\caption{Predicted numbers of $4 \gamma \; e^+ e^-$ events corresponding to
$3S \gto 2P \gto 1D \gto 1P \gto 1S \to e^+ e^-$ or $3S \gto 2P \gto 2S \gto 1P
\gto 1S \to e^+ e^-$ per $10^6~\Upsilon(3S)$
decays.  From Ref. \cite{gr-dw}. where photon energies are quoted.}
\begin{center}
\begin{tabular}{c c c c r} \hline
$2^3P_J$ state & Next state & $1^3P_J$   & Events \\
		&		& state		&	\\
\hline
$2^3P_2$  & $1^3D_3$ & $1^3P_2$ & 7.8 \\
               & $1^3D_2$ & $1^3P_2$ & 0.3 \\
               &                & $1^3P_1$ & 2.7 \\
               & $1^3D_1$ & $1^3P_2$ & 0.0 \\
               &                & $1^3P_1$ & 0.1 \\
               &                & $1^3P_0$ & 0.0 \\ \cline{2-4}
               & $2^3S_1$ & $1^3P_2$ & 4.1 \\
               &                & $1^3P_1$ & 8.8 \\
               &                & $1^3P_0$ & 0.4 \\ \hline
$2^3P_1$ & $1^3D_2$  & $1^3P_2$ & 2.5 \\
               &                & $1^3P_1$ & 20.1 \\
               & $1^3D_1$ & $1^3P_2$ & 0.1 \\
               &                & $1^3P_1$ & 3.3 \\
               &                & $1^3P_0$ & 0.4 \\ \cline{2-4}
               & $2^3S_1$ & $1^3P_2$ & 7.5 \\
               &                & $1^3P_1$ & 15.9 \\
               &                & $1^3P_0$ & 0.7 \\ \hline
$2^3P_0$ & $1^3D_1$  & $1^3P_2$ & 0.0 \\
               &                & $1^3P_1$ & 0.3 \\
               &                & $1^3P_0$ & 0.0 \\ \cline{2-4}
               & $2^3S_1$ & $1^3P_2$ & 0.3 \\
               &                & $1^3P_1$ & 0.7 \\
               &                & $1^3P_0$ & 0.0 \\ \hline
\end{tabular}
\end{center}
\end{table}

\subsection{$\eta_b(nS)$ States}

$\eta_b(nS)$ can be produced via magnetic dipole (M1) transitions
$\Upsilon(nS)\to \eta_b(n' S) +\gamma$ \cite{gr-etab}.  The rates 
for magnetic dipole transitions in quarkonium ($Q \bar Q$) bound
states are given in the nonrelativistic approximation by 
\begin{equation}
\Gamma(^3S_1 \to ^1S_0 + \gamma) = \frac{4}{3} \alpha \frac{e_Q^2}{m_Q^2}
|\bra{f} j_0(kr/2) \ket{i}|^2 \omega^3
\end{equation}
where $\alpha$ is the fine-structure constant, $e_Q$ is the
quark charge in units of $|e|$ ($-1/3$ for $Q=b$), $m_Q$ is the quark
mass (which we take equal to 4.8 GeV/$c^2$)
and $k\; (\omega)$ is the emitted photon's momentum (energy).  
The available phase 
space for allowed transitions (the principal quantum number 
is unchanged) is small leading to a small partial width.  In 
contrast, the hindered transitions (the principal quantum 
number changes) have large available phase space.  
In the nonrelativistic 
limit the overlap integral for hindered transitions is zero due to the 
orthogonality of the initial and final wavefunctions.  However, 
relativistic corrections from the hyperfine interaction leads to
differences in the $^3S_1$ and $^1S_0$ wavefunctions  resulting in a 
non-zero overlap.  The branching ratios for 
$\Upsilon(nS)\to \eta_b(n'S) \gamma $ are given in Table 2.  
Note that the radiative widths are quite sensitive 
to model dependent details. From 
these predictions we expect $\eta_b$'s to be produced at a substantial 
rate.  Another possible production channel for $\eta_b$'s is the 
process $\Upsilon(3S)\to h_b (^1P_1) \pi \pi \to \eta_b + \gamma +\pi 
\pi$ \cite{ky81}.  
The BR for  $\Upsilon(3S)\to h_b (^1P_1) \pi \pi$ is expected 
to be 0.1-1~\% while the BR for  $h_b (^1P_1) \to \eta_b + \gamma$ is 
estimated to be $\sim 50\%$.

\begin{table}
\caption{Predictions for branching ratios of M1
transitions between $n^3S_1$ and $n'^1S_0$ $b \bar b$ levels taking into
account relativistic corrections. From Ref. \cite{gr-etab} using 
wavefunctions from Ref. \cite{gi85}.}
\begin{center}
\begin{tabular}{l c c } \hline
          & Transition & \br $(10^{-4})$ \\
\hline
$\Upsilon (3S)$ 		& $\to 3^1S_0$ & 0.10 \\
$(\Gamma_{tot}=52.5$~keV)	& $\to 2^1S_0$ & 4.7 \\
				& $\to 1^1S_0$ & 25 \\
\hline
$\Upsilon (2S)$ 		& $\to 2^1S_0$ & 0.21 \\
$(\Gamma_{tot}=44$~keV)		& $\to 1^1S_0$ & 13 \\
\hline
$\Upsilon (1S)$ 		& $\to 1^1S_0$ & 2.2 \\
$(\Gamma_{tot}=26.3$~keV)	& 	       &  \\
\hline
\end{tabular}
\end{center}
\end{table}

\subsection{Production of $^1P_1$ States}

We consider two types of cascades to produce the $^1P_1$ $c\bar{c}$ 
and $b\bar{b}$ states \cite{gr-hb}.  
The first starts with an $\Upsilon(3S)$ or $\psi(2S)$ produced 
in $e^+e^-$ annihilation which then underdoes an M1 transition to the 
$\eta(2S)$ state followed by an E1 transition to the $h(^1P_1)$ state 
and a final E1 transition to the $\eta(1S)$ state:
\begin{equation}
\Upsilon(3S)\to \eta_b(2S) +\gamma \to h_b +\gamma\gamma \to \eta_b 
+\gamma\gamma\gamma
\end{equation}
\begin{equation}
\psi(2S)\to \eta_c(2S) +\gamma \to h_c +\gamma\gamma \to \eta_c 
+\gamma\gamma\gamma
\end{equation}
The second cascade consists of the single pion emission 
transition from the initial $^3S_1$ state followed by an E1 transition 
to the $\eta(1S)$:
\begin{equation}
\Upsilon(3S)\to h_b +\pi \to \eta_b +\gamma +\pi
\end{equation}
\begin{equation}
\psi(2S)\to h_c +\pi \to \eta_c +\gamma +\pi
\end{equation}
To determine if either of these cascades is  promising 
we need the appropriate branching ratios and hence the partial 
widths.  In the quark model they are given by:
\begin{equation} 
\Gamma(^1S_0   \to ^1P_1 + \gamma)
=  \frac{4}{3} \alpha \; e_Q^2 \; 
\omega^3 \; |\langle ^1P_1 | r | ^1S_0 \rangle |^2 
\end{equation}
so that $\Gamma[\eta_b(2S)\to h_b(1P)+\gamma] =2.3$~keV. 
In addition we need the hadronic widths for the $\eta_b(2S)$. 
There are considerable theoretical 
uncertainties which to a large extent can be reduced by relating 
ratios of theoretical expressions to observed hadronic widths.  For 
the $\eta_b(2^1S_0)$ we take
\begin{equation}
\Gamma[^1S_0\to gg] 
= {{27 \pi}\over {5(\pi^2-9)\alpha_s}} 
\times \Gamma[^3S_1 \to ggg]
\end{equation}
where 1st order QCD corrections are included in our numerical results 
but are 
not explicitly shown.  This gives 
$\Gamma [\eta_b(2^1S_0) \to gg]= 4.1 \pm 0.7$~MeV.
Combining the resulting  branching ratios for 
\br$[\Upsilon(3S)\to \eta_b' + \gamma]$ and \br$[\eta_b'\to h_b 
+\gamma]$ gives 
\br$[\Upsilon(3S)\to \eta_b' \gamma \to h_b 
\gamma\gamma]=2.6\times 10^{-7}$.  This would result in only 0.3 
events per $10^6 \; \Upsilon(3S)$'s.  A similar exercise yields 
$\br[\psi(2S)\to \eta_c' \gamma \to h_c \gamma\gamma]=10^{-6}$ or 1 
event per $10^6 \; \psi'$'s.  

A more promising approach utilizes the hadronic decay $\Upsilon(3S)\to 
\pi 1^1P_1$ which is estimated to have a BR of around 0.1\%
\cite{voloshin-hc}.  Again we 
need to estimate the radiative and hadronic widths of the $^1P_1$ 
states:
\begin{equation}
\Gamma[^1P_1 \to ^1S_0 + \gamma] 
= \frac{4}{9} \alpha \; e_Q^2 \; 
\omega^3 \; |\langle ^1S_0 | r | ^1P_1 \rangle |^2
\end{equation}
giving $\Gamma[h_b(1P) \to\eta_b(1S)+\gamma]=37$~keV and
\begin{equation}
\Gamma[^1P_1 \to ggg] 
= {{5}\over {2n_f}} \times \Gamma[^3P_1 \to q\bar{q}g]
\end{equation}
giving $\Gamma[h_b(1P) \to ggg] = 50.8$~keV.
Combining the branching ratios in this decay chain yields
\begin{equation}
\hbox{\br}[\Upsilon(3S)\to  1^1P_1 +\pi \to 1^1S_0 +\gamma \pi]= 4 \times 10^{-4}
\end{equation}
which would yield 400 events per $10^6$ $\Upsilon(3S)$'s.  Similarly 
one obtains 
\begin{equation}
\hbox{\br}[\psi(2S)\to  1^1P_1 + \pi \to 1^1S_0 +\gamma \pi]= 3.8 \times 10^{-4}
\end{equation}
In both cases it appears that the $1^1P_1$ should be produced in 
sufficient numbers to be observed.  

%\begin{figure}[htb]
%\vspace{9pt}
%\centerline{\includegraphics[width=15pc]{m1.eps}}
%\caption{Radiative transitions in the $b\bar{b}$ system.  The dashed 
%lines represent $M1$ transitions, the solid lines $E1$ transitions and 
%the dotted lines single $\pi^0$ emission.  The transitions are labelled with 
%their partial widths given in keV.}
%\end{figure}

There is also the possibility that charmonium states can 
be observed in $B$ decay.  This was recently highlighted by the Belle 
observation of the $\eta_c(2S)$ in $B\to \eta_c(2S) K \to K K_s K^- 
\pi^+$ \cite{belle02}.  Belle had previously reported the observation of 
the $\chi_{c0}$, $\chi_{c2}$ \cite{belle-chi} 
and the $\chi_{c1}$ had been observed by 
both the BaBar and Belle collaborations\cite{babar-chi,belle-chi}.  
It had long been proposed 
that the $h_c$ could be seen in $B$ decays \cite{btoc-o}
and the recent Belle result 
has led to a renewal of interest in this process \cite{btoc-n}. 

\section{S-WAVES BELOW 1 GEV}

The subject of the S-waves below 1~GeV garners some of the most vigorous 
discussion in hadron spectroscopy \cite{ct02}.  Thus,  the discussion below 
reflects my impression of this subject which may not be held by all 
workers in this field.  Despite this note of caution I believe that 
there is a growing consensus on the general features of S-wave 
scattering in this kinematic region, if not on all specific details.

The phenomenology of the $0^{++}$ sector reflects the many different 
components of the Fock space: conventional $q\bar{q}$ mesons, 
$q\bar{q}q\bar{q}$ states, glueballs, meson-meson molecules, and 
threshold effects.  It is clearly a challenge to understand this 
sector.  One starts by analyzing the raw data to produce phase shift 
plots and 
then one attempts to intrepret these results in terms of underlying 
models.

Much of the information on low energy S-wave $\pi\pi$ scattering comes 
from measuring $\pi^- p \to \pi^+\pi^- n$  at CERN \cite{cern}
and $\pi^- p \to \pi^0\pi^0 n$  at BNL \cite{e852-01}.  
Unfortunately the experiments provide fewer 
observables than are needed to unambiguously describe the partial 
waves.  One must therefore make assumptions in the PWA.  In particular 
the role of nucleon spin is ignored and the role of the $a_1$ exchange 
amplitude is generally neglected assuming the dominance of pion 
exchange.  These assumptions result in 
ambiguities in the extraction of phase shifts from the PWA.

Kaminski, Lesniak, and Rybicki studied this problem \cite{klr01}  
constraining the allowed solutions by imposing unitarity and using the 
non-observation of inelastic scattering below the $f_0(980)$.  They 
further related the $\pi^+\pi^-$ to the $\pi^0\pi^0$ amplitudes to 
obtain $\pi\pi$ phase shifts.  Including these ingredients they 
presented what they believe to be a unique physical solution.

%\begin{figure}[htb]
%\vspace{9pt}
%\centerline{\epsfig{file=klr-f7.eps,width=15pc,clip=}}
%\caption{Comparison of present phase shifts with....}
%\end{figure}

Furthermore, by separating the pseudoscalar from the pseudovector 
exchange amplitudes they obtain the cross sections shown in Fig. 1.  
The conclusion is that the $a_1$ contributions are significantly 
different from zero.  This is an important point which should be kept 
in mind in future experiments.  High statistics enables the 
differentiation between different t-channel exchange particles which 
results in different couplings to final states.  It is a very useful 
bit of information in understanding the properties of the produced excited 
mesons.

\begin{figure}[htb]
%\vspace{9pt}
\centerline{\epsfig{file=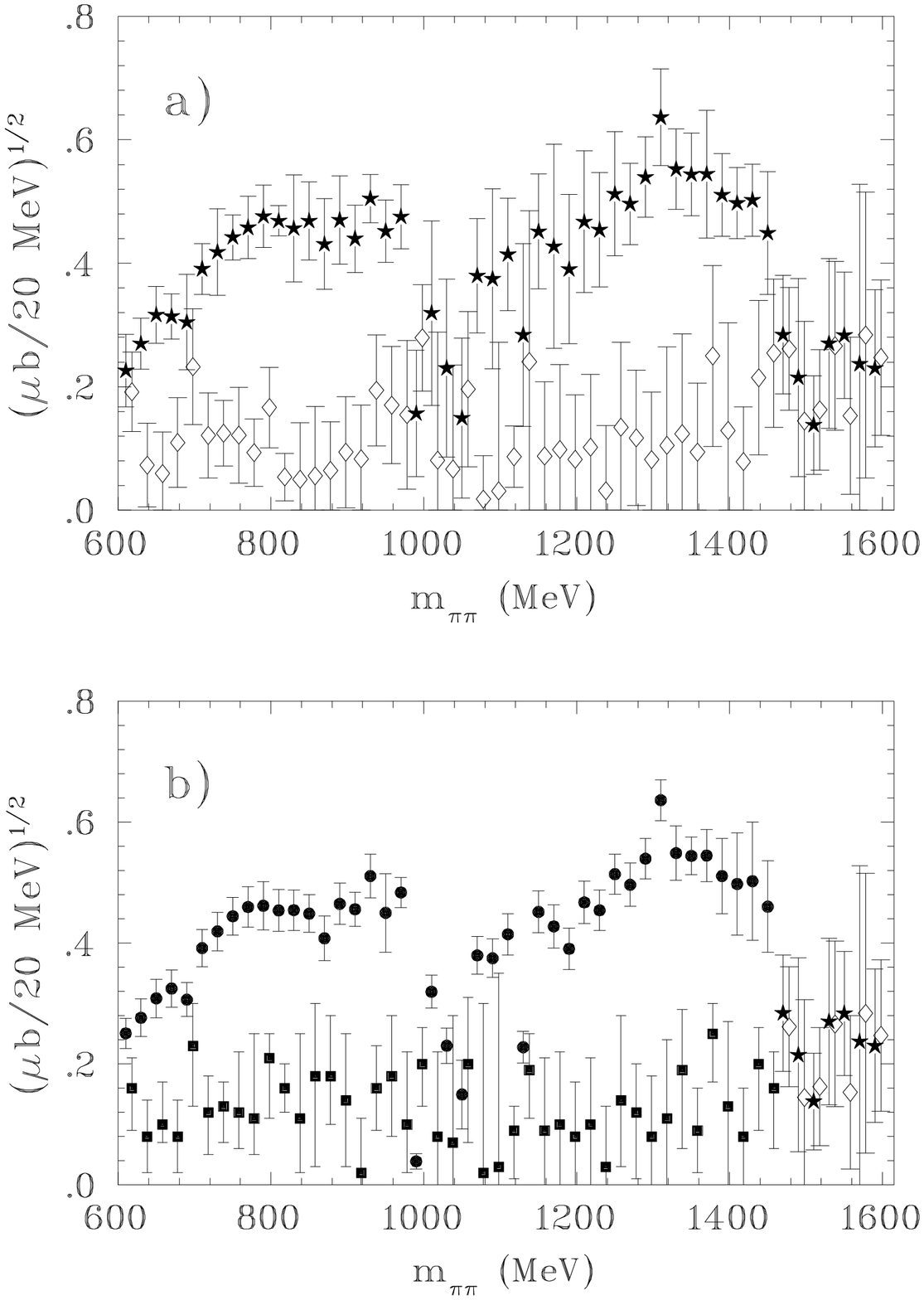,width=16pc,clip=}}
\caption{Moduli of the pseudoscalar (circles) and pseudovector 
amplitudes (squares) for the $\pi\pi$ scattering
``down-flat'' solution of Ref. \cite{klr01}.}
\end{figure}

Given this information the next question is how to describe these 
phase shifts in terms of true resonances and from the dynamics between 
$K\bar{K}$, $K\pi$ and $\pi\pi$.  This was discussed in contributions 
by Van Beveren and 
Rupp \cite{vr02} and by Furman and Lesniak \cite{fl02}.  
It is clear that one must look at the meson-meson 
scattering and extract the phase shifts not the resonance positions.  
To do this requires a multichannel approach which includes resonances 
(from constituent $q\bar{q}$ channels) and meson-meson interactions.  
An interesting and important consequence is that the pole parameters 
are dependent on the coupling strengths.

As a concrete example of this approach I show results for the coupled 
channel calculation of Furman and Lesniak \cite{fl02}.  They 
constructed a coupled channel of two $a_0$ resonances decaying into 
the $\pi\eta$ and $K\bar{K}$ mesons using the coupled channel
Lippman-Schwinger equation:
%\begin{equation}
\begin{eqnarray}
\bra{p}T\ket{q}& & =  \bra{p}V\ket{q} \nonumber \\
& &  + \int {{d^3s}\over{(2\pi)^3}}
\bra{p}V\ket{s}\bra{s}G\ket{s}\bra{s}T\ket{q}
\end{eqnarray}
%\end{equation}
where $T$, $V$, and $G$ are $2\times 2$ matrices with $G$ the matrix 
of propagators.  
The resulting Argand diagrams 
are shown in Fig. 2.  These plots show significantly 
different widths in the two channels which agrees with E852 and 
Crystal Barrel observations.  The $I=0$ sector was also studied in 
this approach.

\begin{figure}[htb]
%\vspace{9pt}
\centerline{\epsfig{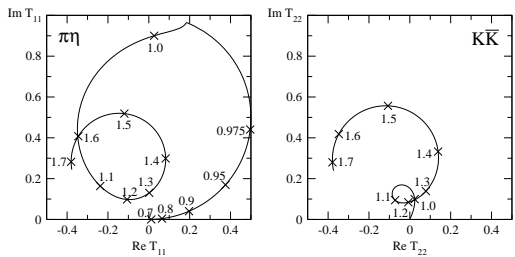}}
\caption{Argand diagrams for $\pi\eta$ and $K\bar{K}$ elastic 
scattering amplitudes.  Numbers denote effective mass in units of 
GeV.  From Ref. \cite{fl02}.}
\end{figure}

What these multichannel approaches by Van Beveren and Rupp \cite{vr02}
and by Furman and Lesniak \cite{fl02} and by others show is that one 
can obtain a good description of the low S-wave spectrum in this 
approach.  My impression is that workers in this field have come to 
a consensus on the general description of S-wave scattering although 
there continues to be considerable debate on the details and 
interpretation of what these results mean.

\section{EXOTIC HYBRID MESONS}

The most important qualitative question in ``soft QCD'' is if 
and how glue manifests itself in the hadron spectrum.  Lattice QCD and 
hadron models predict glueballs, objects with no constituent quark 
content, and hybrids, hadrons with quark content and an explicit gluon 
degree of freedom.  It is expected that the most definitive evidence 
for a non $q\bar{q}$ state is to discover a state with $J^{PC}$ 
quantum numbers inconsistent with the quark model, so called exotics.  
Because the lowest lying exotic glueballs are quite massive this 
directs the search to exotic hybrid mesons.

The emerging picture of hybrid mesons is that of transverse 
excitations of the flux tube.  For the light quarks this leads to the 
lowest mass hybrids of mass $\sim1.9$~GeV with doubly degenerate 
quantum numbers.  In particular, states are predicted with
$J^{PC}=0^{+-}$, $1^{-+}$, and $2^{+-}$ which 
are inconsistent with the quark model and would therefore 
provide the smoking gun for new hadronic physics.
There have been a number of sightings of such states \cite{gn99}.  For 
example the Brookhaven E852 experiment has seen evidence for the 
$\hat{\rho}(1^{-+})$ exotic in the reaction $\pi^- p \to 
\pi^+\pi^-\pi^- p$ \cite{e852-98}.  

Here I briefly mention an ambitious proposal to search for 
and map out exotic gluonic excitations at the Jefferson Laboratory in 
Virginia.  The proposal takes advantage of the fact that exotic hybrid 
mesons have their quark spins aligned in an $S=1$ state.  
In standard hadroproduction the 
incoming beam consists of $\pi$ mesons whose quark spins are in an $S=0$ 
state.  Since exotic hybrids have quark spins 
in an $S=1$ state it is believed that hadroproduction of exotic hybrids 
is suppressed.  In contrast, via vector meson dominance, a photon beam 
consists of quarks in an $S=1$ state so it is believed that
exotic hybrids are more readily produced in diffractive photoproduction
by ``plucking'' the gluon flux tube.  To this end there is a proposal to 
upgrade CEBAF to 12 GeV and produce photons via coherent 
bremsstrahlung with a spectrum that peaks at 
9~GeV.  This project would add considerably to our knowledge of hybrid 
mesons and our understanding of confinement.

\section{SUMMARY}

Over the last while, there have been significant developments in heavy 
quarkonium physics.  We expect this progress to continue with a high 
likelihood that the CESR collaboration should find evidence for the 
$2^1S_0$ $1^1S_0$, $1^1P_1$ $b\bar{b}$ states.  It is also possible 
that they will find evidence for the $3^1S_0$, $1^3D_1$ and $1^3D_3$ 
states.  At the same time we expect that in the near future the
various B-factories will discover some of the missing charmonium 
states in B-decay.  Beyond this, the CLEO-c/CESR-c project would 
provide an additional avenue to study the charmonium system.
Taken together, this would represent the largest advances in 
heavy quarkonium spectroscopy in two decades.  These results would 
provide an important benchmark against which to measure the results of 
lattice QCD and the various models in the literature.

The situation with S-wave scattering at low energies is converging to 
a consensus on the overall details.  Good agreement between theory and 
experiment is achieved by describing low energy scattering in the 
framework of a multichannel quark model.  Nevertheless, there is still 
vigorous debate about the details and the interpretation.

Finally, we mention hybrid meson production as an important tool in 
understanding QCD in the low $Q^2$ regime and the nature of 
confinement.  In the near future we look forward to results from,
for example, the 
COMPASS experiment at CERN and photoproduction results at HERA.  In 
the longer term we expect that the GlueX experiment at Jefferson Lab 
will provide a major leap forward in our understanding of gluon 
dynamics in the  regime of Soft QCD.

\end{document}